\documentclass[oldversion]{aa}

\usepackage{epsfig,graphicx}
\usepackage{natbib}
\begin{document}

\title{Distribution of the molecular absorption in front of the quasar B0218+357}

\author{Muller S. \inst{1} \and Gu\'elin M. \inst{2}
\and Combes F. \inst{3} \and Wiklind T. \inst{4}}

\offprints{muller@asiaa.sinica.edu.tw}
\institute{
Academia Sinica Institute of Astronomy and Astrophysics (ASIAA), P.O. Box 23--141, Taipei, 106 Taiwan
\and
Institut de Radio Astronomie Millim\'etrique (IRAM), 300 rue de la piscine, F--38406 St Martin d'H\`eres, France
\and
Observatoire de Paris, LERMA, 61 av. de l'Observatoire, F--75014 Paris, France
\and
ESA Space Telescope Division, STScI, 3700 San Martin Drive, Baltimore, MD 21218}

\date {Received 16 February 2007 / Accepted 3 April 2007}
%\thesaurus{}

\titlerunning{Distribution of the molecular absorption toward B0218+357}
\authorrunning{Muller et al. 2007}

\abstract{The line of sight to the quasar B0218+357, one of the most studied lensed systems,
intercepts a $z = 0.68$ spiral galaxy, which splits its image into two main components A and B,
separated by ca. 0.3$\arcsec$, and gives rise to molecular absorption. Although the main absorption
component has been shown to arise in front of image A, it is not established whether some
absorption from other velocity components is also occuring in front of image B.
To tackle this question, we have observed the
HCO$^+$(2-1) absorption line during the commissioning phase of the new very extended configuration
of the Plateau de Bure Interferometer, in order to trace the position of the absorption as a function
of frequency. Visibility fitting of the self-calibrated data allowed us to achieve position
accuracy between $\sim$ 12 and 80 mas per velocity component. Our results clearly demonstrate that
all the different velocity components of the HCO$^+$(2-1) absorption arise in front of the south-west
image A of the quasar. We estimate a flux ratio $f_A/f_B = 4.2 _{-1.0} ^{+1.8}$ at 106 GHz.

}

\maketitle

\section{Introduction}

Molecular absorption at intermediate redshifts has been detected only in a few objects
({\em e.g.} \citealt{wik95,wik96,wik97,wik98,kan05}).
Among these, three absorption systems are
caused by a galaxy lying on the line-of-sight of a quasar and acting as a gravitational
lens. Two systems, one in front of PKS1830$-$211 (at $z=0.89$), the other in front of B0218+357 (at $z=0.68$) are
detected at millimeter wavelengths in the lines of several molecules, including HCO$^+$,
HCN and H$_2$CO.

The light rays associated with the quasar PKS1830$-$211 form two bright images that probe
different regions of the intervening galaxy. This provides information on the latter's kinematics,
its mass, as well as on the physical and chemical conditions in its interstellar medium
(\citealt{wik98,mul06}). 
These two papers showed that the two $z \sim 0.89$ absorption components of PKS1830$-$211 are
associated each with one of the two gravitational images.

%%%\citet{wik98} have shown that the two $z \sim 0.89$ absorption components of PKS1830$-$211 are associated each with one of the two gravitational images (see also \citealt{mul06}).
%%%These authors estimate a dynamical mass of 6 -- 9 $\times$ 10$^{10}$ M$_\odot$ within a radius of 3 kpc in the intervening galaxy.

The second mm-absorption system occurs in the line of sight to B0218+357 and is caused by a
galaxy at a redshift of $z = 0.68466$ (\citealt{bro93,car93,wik95}). Two bright images of the
quasar, hereafter referred to as A (to the SW) and B (to the NE), have been
resolved at radio cm wavelengths (\citealt{ode92,pat93}). The distance AB between the two
images is $\sim$ 0.3$\arcsec$, the smallest angular separation among the known
galaxy-mass lenses. The flux ratio is $f_A/f_B$ $\sim$ 3 between 5 GHz and 22 GHz. Image
B lies in the center of an Einstein ring whose diameter is similar to the distance
AB. Each image reveals intricate sub-structures at very high angular resolution
(\citealt{pat95, big03}).  The constraints provided by the complex image pattern and by the
time variability of the background source flux, make of B0218+357 one of the best objects
to measure H$_0$ at intermediate redshifts (\citealt{big03,wuc04,yor05}).
Deep ACS/HST observations of B0218+357 by \citet{yor05} reveal that the lensing object is a
spiral galaxy seen nearly face-on and whose center lies close to image B, the center of the
Einstein ring. According to the lensing model proposed by \citet{wuc04}, image A is located
at about 2 kpc from the center of the lensing mass distribution, and image B at 0.4 kpc. HI
absorption is detected over a velocity width of about 100 km~s$^{-1}$ (\citealt{car93, kan03}),
in front of image A (\citealt{car00}), although some absorption may also occurs in front of the
Einstein ring (\citealt{kan03}). H$_2$CO ($2_{12}-2_{11}$) absorption has been observed by
\citet{men96} towards image A, with a total width of $\sim$ 12 km~s$^{-1}$, much narrower
than the HI profile. Because of limited signal-to-noise ratio, they could however not
exclude the possibility of absorption in front of B.

Fig.\ref{spec} shows the HCO$^+$(1-2) absorption profile towards B0218+357 observed with a
high sensitivity and velocity resolution (Muller et al., in prep.). Like the HCN (1-2)
profile, it shows at least four velocity components over a
width of 25 km~s$^{-1}$, much broader than that of H$_2$CO ($2_{12}-2_{11}$). The component
with the deepest absorption matches in width and velocity the H$_2$CO ($2_{12}-2_{11}$)
absorption profile and most probably originates like this latter, in front of image A.
The other three components, however, had not been observed so far with a high angular
resolution and their location, in front of A, B, or of the Einstein ring remained unknown.
Yet, the knowledge of this location is essential for the interpretation of the molecular
absorption data in terms of line opacities, molecular column densities and gas kinematics.

In this letter, we present observations of the HCO$^+$(1-2) line toward B0218+357, obtained
with the new, extended configuration of the Plateau de Bure Interferometer, which allow us
to trace the position of all four absorption components.

\section{Observations and data analysis}

\begin{figure}[th] %\resizebox{\hsize}{!}
\includegraphics[width=7.5cm]{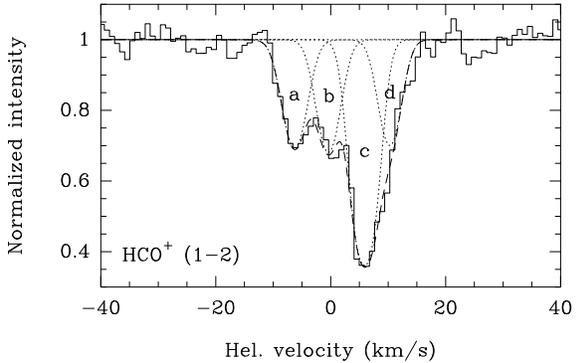}
\caption{Spectrum of the HCO$^+$(1-2) absorption towards B0218+357 as observed with the
PdBI in a compact configuration. Gaussian fits of the different velocity components a, b,
c, d are overlaid as dashed lines. 
%The velocity resolution is 0.9 km~s$^{-1}$.
} \label{spec}
\end{figure}

In the frame of a survey of molecular absorption lines towards B0218+357,
we have observed the HCO$^+$(1-2) line in June 2005 and July 2006, with 
a compact configuration of the IRAM Plateau de Bure Interferometer (PdBI).  These observations were
self-calibrated on the continuum source, which was not resolved with the $\sim 4''$ FWHP
synthetized beam. The resulting average HCO$^+$(1-2) spectrum is shown in Fig.\ref{spec}.
The absorption extends over $\sim$ 25 km~s$^{-1}$ and can be decomposed into four Gaussian
velocity components, with FWHM of 4 -- 5 km~s$^{-1}$.

We have further observed on November 8$^{\rm th}$ 2005 the HCO$^+$(1-2) line with an
extended configuration (maximum projected baseline B$_{\rm max}$ of 385 m) and on November 9$^{\rm th}$ 2005 with the
new, very extended configuration of the PdBI (B$_{\rm max}$=650  m). We emphasize that these
observations were the first science data acquired with the new extended configuration and
that they were performed during its commissioning phase. In particular, the instrument
baselines were not yet properly calibrated, the on-source integration time was limited 
%(see Table \ref{obs})
(110 min in each configuration), and the weather not at its best (rms phase deviation prior to
self-calibration of $\sim$ 50$\degr$ on the longest baselines). Therefore, the data needed to be
self-calibrated on the continuum source, which was half-resolved in the EW and NEW
directions. 

As in the June and July sessions, we observed simultaneously a 540 MHz-wide
frequency band centered at 105.690 GHz, consisting of two overlapping 320 MHz-wide
continuum sub-bands (L03 and L04) with a channel spacing of 2.5 MHz (7.1 km~s$^{-1}$), and a 80 MHz-wide
sub-band (L01) centered on the redshifted HCO$^+$(1-2) line frequency $\nu$=105.882 GHz,
with a channel spacing of 0.31 MHz (0.9 km~s$^{-1}$).
The high frequency edge of the continuum band was dropped and its bandwidth restricted to
440 MHz in order to avoid contamination by the HCO$^+$(1-2) line.

The method used to self-calibrate on the half-resolved continuum source was similar to the
one used by \citet{mul06}, for localizing the absorption components in front of
PKS1830$-$211. The data from November 8$^{\rm th}$ and 9$^{\rm th}$ were calibrated
separately, using the GILDAS/CLIC software, with the following procedure: In a first step,
the radio frequency (RF) bandpass of L03 and L04 was calibrated directly on the B0218+357 continuum signal.
%L03 and L04
%%%, which define a 440 MHz-wide band not covering the HCO$^+$ line, 
%were calibrated in RF
%directly on the B0218+357 continuum signal.
The main purpose of using these
absorption-free bands was to correct for short term phase and
intensity variations linked to atmosphere fluctuations by
self-calibrating on the continuum.
%{\bf using high order polynomials (e.g. 6 and 18 for the amplitude and phase respectively).
L01 was calibrated in RF separately, using low order polynomials
(first degree in amplitude and third degree in phase). The narrowness
of the absorption line ($<$ 10 MHz) with respect to the L01 bandwidth
(80 MHz) ensured that the calibration was not badly affected by the
absorption features.
%We come back to this point further down in this Section.
%The narrowness of the absorption components ($<$ 10 MHz) allowed us to use high order
%polynomials (e.g. 6 and 18 for the amplitude and phase respectively) at this stage.
Note that we choose a baseline-based RF calibration,
rather than an antenna-based, as the signal intensities for the longest baselines were
severely affected by atmospheric phase decorrelation.

Next, the continuum sub-bands were
self-calibrated in amplitude and phase by calculating the complex gains corresponding to
a system of two point-like sources separated by $\Delta$R.A. = 309
mas, $\Delta$Dec. = 128 mas, and with a flux ratio of 4.2 (a justification of
theses values will be given in \S3, and/or can be found in the reviews by
\citealt{wuc04,mit06a,mit06b}). 
The total continuum intensity ($\sim$ 0.4 Jy) was normalized to unity.
The gains calculated for the continuum sub-bands were then applied
to the visibilities of the L01 sub-band channels. Finally, the continuum visibilities, as
derived from the double-source model, were subtracted from the calibrated L01 visibilities.

The calibrated L01 datasets from November 8$^{\rm th}$ and 9$^{\rm th}$ were then combined.
The visibilities corresponding to uv radii larger than 200 m were fitted, channel by
channel, with a single point-like source representing the position and strength of the
absorption signal. Visibilities with shorter uv radii were discarded, as they bring little
information on the signal position. The fitted source intensities showed a small offset
($\sim$ 10\% of the total continuum)
with respect to zero (the continuum level as defined from the continuum sub-bands) --
probably the result of residual IF and/or RF bandpass calibration errors. This was taken care
of by averaging the visibilities of the L01 channels that are free of any absorption and by
subtracting the so-computed residual continuum visibility from all L01 channels.
The fit of the point-like absorption sources was then repeated.

\begin{figure}[h] %\resizebox{\hsize}{!}
\includegraphics[width=7.5cm]{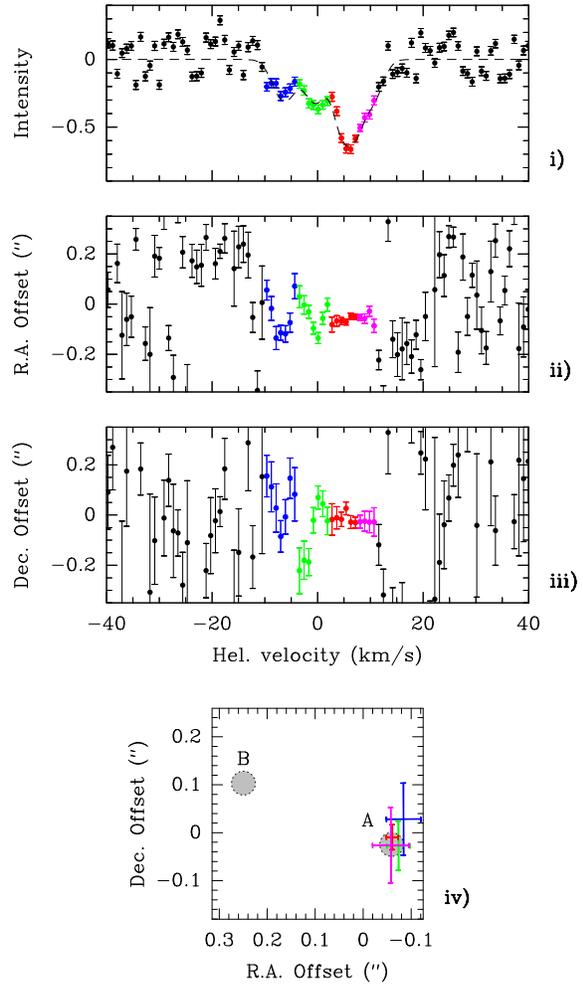}
\caption{Relative intensity and position of the centroid of the HCO$^+$(1-2) absorption 
%in the HCO$^+$(1-2) line, as a function of velocity. The intensities and positions are 
as derived from the fit of a point like-source to the self-calibrated visibilities, for each velocity channel. 
{\bf \em i)} The intensity scale is zero at the continuum level and at $-$1 for full
absorption.
The profile of the absorption, as fitted in Fig.\ref{spec}, is overlaid as a dashed line.
{\bf \em ii, iii)} Position offsets relative to the centroid of the continuum.
{\bf \em iv)} Position of the four Gaussian velocity components, compared with the locations of the
two lensed images of the quasar (grey disks, not to scale).} 
\label{uvfit}
\end{figure}

Fig.\ref{uvfit} shows the result of the final fits. The upper plot ({\em i}) shows the
absorption profile as a function of velocity, while the two middle plots ({\em ii} and
{\em iii}) show the position of the absorption (or, more exactly, the position of the
centroid of the absorption) as a function of velocity. The removal of the continuum insures
that zero intensity (0) and zero offset (0,0) effectively correspond to the intensity and
position of the centroid of the continuum sources. The source position accuracy in the
individual velocity channels was calculated by the GILDAS fitting routine UVFIT. It can be
crudely expressed as $\sim$ $\theta _{\rm beam}$ / 2 SNR, where $\theta _{\rm beam}$ is
the FWHP of the synthesized beam ($\simeq 1.2 \times 0.7\arcsec$, P.A. 176$\degr$)
and SNR, the signal-to-noise ratio
on the source intensity. The noise level, per 0.9 km~s$^{-1}$ velocity channel, is 3.4\% of
the continuum intensity.
For all channels with absorption signal ($-$10 $<$ V$_{\rm HEL}$ $<$ +12 km~s$^{-1}$), the
fitted phases indicates that, within the uncertainties, the absorption arises from a small
size region that coincides in position with image A. We further averaged the position
offsets per velocity components, with weights equal to the square of
the intensity of the Gaussian profiles.
The average positions are given in Table \ref{posfit} and Fig.\ref{uvfit}{\em iv}.

\section{Discussion}

Our data show that all the molecular absorption components (from V$_{\rm HEL}$ = $-$10 to
+12 km~s$^{-1}$) originate in front of image A. This is not surprising since
% {\em i)} there is a large differential Faraday rotation ($\sim$ 900 rad~m$^{-2}$,
%\citealt{pat93}) presumably due to the different magneto-ionic medium between locations A and B
%in the lensing galaxy, and {\em ii)}
image A appears more obscured in the visible than image B (\citealt{gru95}). As was mentioned,
\citet{men96} have already shown that the V$_{\rm HEL}$ $\sim$ 5 km~s$^{-1}$ H$_2$CO
absorption component arose in front of image A.

The background radio source presents a core-jet morphology that is reflected in both lensed
images in the form of sub-structures. Those consist mainly in two main components, labelled A1
and A2 (and B1 and B2), which are separated by $\sim$ 1.4 mas (\citealt{pat95}). High
sensitivity radio imaging reveals in addition the presence of a weak, relatively extended
(10-15 mas)  knotty jet and of a possible counterjet (\citealt{big03}). Our source positioning
accuracy is not good enough to spatially resolve any of these sub-components. However, the
jet has a steeper spectrum than the core and its emission, which is already weak in the
radio domain should be negligible at millimeter wavelengths. The relevant size of the
background continuum, regarding millimeter absorption, should be $<$ 4 mas$^2$.
Assuming a flat universe, with standard cosmological parameters (H$_0$=70 km~s$^{-1}$Mpc$^{-1}$,
$\Omega_{\rm M}$=0.3, $\Omega_\Lambda$=0.7), the angular size scale is 7.1 kpc/$\arcsec$ at a
redshift $z = 0.68$. The mm continuum emission of image A should therefore have an extent lower
than $\sim$ 30 pc$^2$. This value, corrected from the filling factor, gives an estimate of the size
of the molecular absorbing clouds.

\begin{figure}[h] %\resizebox{\hsize}{!}
\includegraphics[width=6cm]{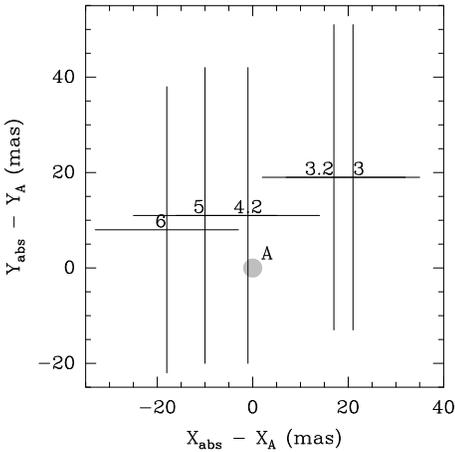}
\caption{Average positions of the HCO$^+$(1-2) absorption features obtained when varying
the flux ratio $f_A/f_B$ (indicated above each position) in the source model used for the self-calibration.} 
\label{ratio}
\end{figure}

It is difficult to determine a {\em direct} value of the flux ratio $f_A/f_B$ from our
current data. Nevertheless, we have repeated the self-calibration procedure described in
\S2, by varying the flux ratio in the source model. The fitted positions of the
absorption, averaged over V$_{\rm HEL}$ = $-$10 to +12 km~s$^{-1}$, were then compared to
the position of image A set by the source model (see Fig.\ref{ratio}). Given the size of
the continuum source, the average position of the absorption is consistent with the
position of image A for $f_A/f_B = 4.2 _{-1.0} ^{+1.8}$. We emphasize at this point that
the average position derived in this way changes by less than $\pm 20$ mas when $f_A/f_B$ varies
from 3 to 6, i.e. by the range of possible $f_A/f_B$ ratios, so that our
conclusion that all the absorption arises in front of A is robust. Similarly, the
self-calibration method that we have used depends little on the continuum source model, in
particular on the distance between A and B, so that the uncertainties on this distance do
not affect our results.

Although indirect, our measurement of the flux ratio $f_A/f_B$ is the first at frequencies
higher than 22 GHz. The value of 4.2 is slightly higher than those measured at 15 and 22 GHz with the
VLA (\citealt{ode92, pat93,big99}), and almost twice higher than those observed around 2 GHz
(\citealt{mit06a}). This is consistent with the model developed by \citet{mit06b},
where image A is obscured at radio frequencies by a HII region associated with the
molecular cloud. We note that the $f_A/f_B$ ratio might be affected by the time variability
of the quasar, due to the time difference $\Delta t$ between the transit times of the light
in A and B. The time delay, however, is relatively short ($\Delta t$ $\sim$ 10 days,
\citealt{big99}), making the chance of a factor of 1.5-2 variation of the quasar intensity
in less than $\Delta t$ rather unlikely.

\begin{table*}[ht]
\caption{Average positions of the different absorption components. For comparison, the positions
of images A and B, as fixed in our continuum source model, are indicated.
Offsets are given relatively to the barycenter of the continuum emission.}
\label{posfit}
\begin{center} \begin{tabular}{cccccc}
\hline
Velocity  & V$_{\rm HEL}$$^\dagger$ & Absorption & $\Delta$V$_{1/2}$$^\dagger$ & R.A. offset & Dec. offset \\
component & (km~s$^{-1}$) & depth$^\dagger$ $^\ddagger$     & (km~s$^{-1}$) & (mas) & (mas) \\
\hline
a       & $-$6.3    & 0.31 & 4.7   & $-$84 $\pm$ 36 & +28 $\pm$ 75 \\
b       & $-$0.3    & 0.32 & 4.5   & $-$73 $\pm$ 24 & $-$28 $\pm$ 51 \\
c       & +5.9      & 0.63 & 4.5   & $-$60 $\pm$ 12 & $-$11 $\pm$ 26 \\
d       & +10.3     & 0.30 & 4.4   & $-$57 $\pm$ 38 & $-$27 $\pm$ 79 \\
a+b+c+d & $-10$ to +12 & & & $-$61 $\pm$ 15 & $-$15 $\pm$ 32 \\
\hline
Image A  &      & & & $-$59        & $-$25 \\
Image B  &      & & & +250          & +103 \\
\hline
\end{tabular} \end{center}
\mbox{\,} \vskip -.3cm
$^\dagger$ from the fit of the spectrum obtained in the compact configuration (cf. Fig.\ref{spec}).\\
$^\ddagger$ with respect to the total continuum intensity.
\end{table*}

Adopting $f_A/f_B$ = 4.2 and assuming a filling factor $f_c$ = 1, we derive a maximum
optical depth $\tau _{\rm max}$ = 1.5 for component c,
%at V$_{\rm HEL}$ $\sim$ 6 km~s$^{-1}$,
i.e. the peak opacity is large. Conversely, a lower limit to the filling factor $f_c$ can
be derived by assuming $\tau _{\rm max} \rightarrow \infty$; this yields $f_c$ $\geq$ 0.77. The H$_2$CO
(\citealt{men96}), NH$_3$ (\citealt{hen05}) and H$_2$O (\citealt{com97}) absorptions are probably
caused by the same cloud as component c. For the other velocity components, the filling
factors and/or the peak opacities must be lower. A filling factor $f_c$ $\simeq$ 0.9 is
consistent with the fact that image A, despite a high column density of absorbing gas, is
strongly attenuated on optical V-band images, with respect to H-band images, but still
visible. It might also naturally explain why the optical distance AB ($\sim$ 317 mas,
\citealt{yor05}) appears lower than in radio ($\sim$ 334 mas). The optical barycenter of
image A should indeed be shifted closer to image B, if the absorption occurs mostly on the
opposite border.

By co-adding all our observations in the compact configuration
of the PdBI at the frequency of the HCO$^+$(1-2) line,
%%%(June 2005 + July 2006)
we do not detect any new absorption
feature, outside the V$_{\rm HEL}$ = $-$10 to +12 km~s$^{-1}$ components just described, over
the range V$_{\rm HEL}$ = $-$200 to +1300 km~s$^{-1}$, and this down to a level of 3$\sigma$
= 3.7\% of the continuum at a velocity resolution of 7.1 km~s$^{-1}$. 
As the intervening galaxy is seen face-on, it is unlikely that the
difference in velocity between positions A and B, located at either sides of the center of
the galaxy, is more than 200 km~s$^{-1}$. HCO$^+$ is known to be easily detectable in
absorption in the Galaxy, even in diffuse molecular clouds (\citealt{luc96}).
Therefore, either the region intercepted by the light path associated with image B and
located at $\sim$ 400 pc from the center of the galaxy, is free or almost free of molecular
gas ($\tau_{\rm HCO+}$ $<$ 0.2), or the filling factor is low ($f_c$ $<$ 0.2).

\section{Conclusion}

Sensitive observations of the quasar B0218+357 in the HCO$^+$(1-2) line, redshifted by $z = 0.68$,
show at least four velocity components in absorption with velocities between V$_{\rm HEL}$ = $-$6
to +10 km~s$^{-1}$ and widths $\simeq$ 5 km~s$^{-1}$. By fitting the visibilities
obtained from new observations with the very extended configuration of the Plateau de Bure
Interferometer, we show unequivocally that all these components arise in front of the SW
gravitational image (A) of the quasar. We see no other HCO$^+$ absorption components over
a range V$_{\rm HEL}$ = $-$200 to +1300 km~s$^{-1}$, that could arise from image A,
from the NE image B, or from the weak Einstein ring visible at radio frequencies. This
implies either a low average column density of HCO$^+$ ($\tau$ $<$ 0.2) in front of B and,
presumably, a low average molecular hydrogen column density, or a low filling factor
($f_c$ $<$ 0.2). We derive a flux ratio
$f_A/f_B = 4.2 _{-1.0} ^{+1.8}$ for the two main components, A and B, at 106 GHz, which is slightly
higher than those derived between 15 and 22 GHz and almost twice larger than observed at frequencies around 2
GHz. This measurement, strictly speaking, applies only to Nov 8 \& 9, 2005, the dates of
our extended configuration observations.

\begin{acknowledgements}
The results presented here are based on observations carried out with the IRAM Plateau de
Bure Interferometer. We would like to thank all the people who have been working on the
extension of the interferometer tracks and, more particularly, the IRAM staff that has
carried out these observations. 
We thank the referee, Prof. Karl Menten, for constructive comments.
IRAM is supported by INSU/CNRS (France), MPG (Germany) and
IGN (Spain).
\end{acknowledgements}

\end{document}